\documentclass[proof]{WileyASNA-v1}
\articletype{EDITOR'S CHOICE}%

\received{:17 April 2019}
\revised{:17 July 2019}
\accepted{:5 August 2019}

\begin{document}

\title{An Oppositeness in the Cosmology: Distribution of the Gamma-Ray Bursts and the Cosmological Principle}


\author{Attila M\'esz\'aros}





\address{\orgdiv{Astronomical Institute, Faculty of Mathematics and Physics}, \orgname{Charles University},
\orgaddress{\state{Czech Republic}}}




\corres{Attila M\'esz\'aros,
Astronomical Institute, Faculty of Mathematics and Physics
Charles University,
Czech Republic
CZ-180 00 Prague 8, Hole\v{s}ovi\v{c}k\'ach 747/2.\\
\email{meszaros@cesnet.cz}}



\abstract{The Cosmological Principle is 
the assumption that the universe is spatially homogeneous and isotropic in the large-scale average. 
In year 1998 the author, together with his two colleagues, has shown that the BATSE's short gamma-ray bursts are not distributed isotropically on the sky. This claim was then followed by other papers confirming both the existence of anisotropies in the angular distribution of bursts and the existence of huge $Gpc$ structures in the spatial distribution. These observational facts are in contradiction with the Cosmological Principle, because the large scale averaging hardly can be provided. The aim of this contribution is to survey these publications.}






\maketitle

\section{Cosmological Principle from the Observational Point of View}\label{sec1}

The Cosmological Principle (in what follows CP) requires that the Universe should be
spatially homogeneous and isotropic on scales larger than the size of any
structure (~\cite{Peebles1993}). On the page 15 of this book it is said:
"...in the large scale average the visible parts of our universe are isotropic and homogeneous". 

The observable part of the Universe has the size of $\sim (10-20) \,Gpc$, 
if one uses the so called "proper-motion distance". The relevant formulae are given, e.g., by
~\citet {Weinberg1972} and ~\citet{Carroll1992}. The exact value depends on the omega-factors and on the Hubble-constant. But, in any case, the
observable part is finite, if the so-called  "proper-motion distance" is used, which is always given by luminosity-distance$/(1+z)$, where $z$ is the redshift.

Trivially, the averaging
should happen far below the $\sim (10-20) \,Gpc$ scale. In other words, there should
exist a transition scale not larger than, say, $\sim 1  \, Gpc$, and above this one no structures should exist.
But, if there were observed structures of scales $\sim 1 \, Gpc$, then CP hardly can hold.

There are several classical observational tests based on statistics to verify the fulfilment of CP. These tests are strongly biased by the absorption at the Galactical plane. Due to the limited size of this contribution for illustration we only mention
few tests here.  ~\citet{Abell1961}  used the method of the counts of the clusters of galaxies in cells. ~\citet{Peebles1980} defined and used the so called auto-correlation function for galaxies. For the  description of the clustering of objects the method of the nearest-neighbourhood on the sphere is well applicable (~\citet{ScottTout1989}, ~\citet{SlechtaMeszaros1997}). In \citet{ScottTout1989} this 
method for the second, third, .. k-th neighbourhood is presented by, too.
There are many other methods, too (Voronoi tessellation, minimal spanning tree, multifractal
spectrum, ...) - survey of these methods with references can be found in ~\citet{Vavrek2008}.
The method of the spherical harmonics on the sky is a well-known method in the theory of cosmological density fields (see p.518 of ~\citet{Peacock1999}). But this method is well usable in the isotropy tests of any objects scattered on the sky, because
in the case of isotropy any spherical harmonic itself should be zero. At the last decades the
numerical methods (cf. Monte Carlo simulations) are broadly used (~\cite{Press2002}).

It is widely accepted that the observations confirm CP and it holds. 
 ~\citet{Yadav2010} means that this transition scale is $\simeq 260h^{-1} \, Mpc$ (i.e., it is below the $\sim 1 \, Gpc$ scale), where $h$ is the Hubble-constant in unit $100 \, km/(sMpc)$.
But, on the other hand, there are publications with observational supports for the structures with $\sim \, Gpc$ sizes. From the earlier ones ~\citet{Abell1961} should be mentioned, because it speaks about the clusters of clusters. In addition,  for example, ~\citet{Collins1973} and ~\citet{Birch1982} speak about a possible
global rotation in the observable part of the Universe. Other observations
(cf., ~\citet{Rudnick2007})
claim the existence of structure with size $\sim 140\, Mpc$, but at redshift around $1$.
A recent publication about the spatial distribution of quasars (cf. ~\citet{Clowes2013})
claims the existence of a structure with a scale $> 1\, Gpc$.

The spatial distribution of the gamma-ray bursts (hereafter GRBs) allows well to test CP, too. In any case, the GRBs should be distributed isotropically on the sky as any other cosmological objects. But, for these tests GRBs are especially useful,
because they are seen in the gamma-band also in the Galactic plane, too, and thus there is no observational bias following from the absorption at the Galactical plane.
In year 1996 it was declared that the isotropic angular distribution
of BATSE's bursts was fulfilled (see ~\citet{Tegmark1996} and the references therein).
It is essential to note that in year 1996 no direct redshifts were known for GRBs, and only indirect statistical estimations existed (see, cf., ~\citet{1996ApJ...466...29M} and the references therein). Hence, these tests are in essence 2D tests on the sky.
After 1997, when the first direct redshift was determined (see ~\citet{1997Natur.387..783C}),
for the limited sample of GRBs, when the redshifts are directly measured, not only the angular distribution can be tested, but also the three-dimensional spatial distribution.

This means that from the observational point of view the testing of the fulfilment of CP by
the spatial distribution of GRBs is not biased by the absorption at the Galactical plane. On the other hand, only a small fraction of GRBs has directly measured redshifts. Therefore,
the tests done in essence on the entire samples can be only 2D tests - the 3D tests are possible, too, but on the samples containing much less objects.

In this review article these observational tests - based on the spatial distribution of GRBs - are surveyed.

\section{Redshifts and Subsamples of GRBs: A Brief Description}\label{sec2}

The first article about the discovery of 16 GRBs was published in 1973  (\cite{Klebesadel1973}).
In period 1973-1990 $\simeq (10-20)$ GRBs were detected annually. .
In years 1990-2000 the BATSE instrument on the Compton Gamma-Ray Observatory increased the number of detected GRBs (annually $\sim\, 300$) and confirmed that there are two types of GRBs (the "short/hard" and the "long/soft" types). It was also confirmed indirectly that both types of GRBs are at cosmological distances (see 
\citet{1986ApJ...308L..43P},
\citet{Meegan1992}, 
\citet{1995ApJ...449....9M}, \citet{1996ApJ...466...29M}, \citet{1996ApJ...470...56H},
\citet{Goldstein2013} and the references therein). In year 2003 it was also shown from the analyses of observations that these two types are astrophysically different phenomena
(~\cite{2003A&A...401..129B}).
This means that both types separately should be distributed isotropically on the sky, if CP holds.

The indirect proof of the cosmological origin followed from the statistical studies of the angular distribution of GRBs - they did not show any concentration toward the Galactical plane (\cite{Meegan1992}, \cite{Tegmark1996}). Hence, the cosmological origin was proclaimed without knowing the direct values of redshifts. But, statistically, even before 1997, 
it was also claimed that the redshifts should be up to $z \sim 20$
(\cite{1996ApJ...466...29M}). It seems that the long GRBs are at higher redshifts than the short ones, but this claim is in doubt (~\citep{2007ApJ...664.1000B}).

At year 1997 a long GRB was detected also at other photon energy bands. The so-called "afterglow" was followed after the discovery of the BeppoSAX
satellite (\cite{Costa1997}). After that, at the coming years, it was observationally confirmed that the long GRBs are connected
to the supernovae (for details see, e.g., \citet{Woosley2006}). For the short ones only in 2013 came the observational support that they are
given by the merging of two neutron stars (black holes) forming macronovae (\cite{Tanvir2013}).
The simultaneous detection of the gravitational wave and macronova (also the term "kilonova" is used) from 17 August 2017 confirmed that the observed gravitational waves and the short GRBs should have common origin (\cite{vonKienlin2017}).

It has to be noted that there are several statistical studies claiming that - beyond the short and long ones - also other subgroups exists for GRBs (for details and other issues see, e.g., ~\citet{Levan2014},
 ~\citet{Ripa2012}, ~\citet{Ripa2016} and the references therein). Summing up these efforts it can be said that there are several statistical supports for the existence different subclasses of GRBs. But, on the other hand, no unambiguous proof exists for the existence of more than two types from the astrophysical point of view. In any case, from the statistical point of view, the isotropy of any subclass should be fulfilled separately, if CP holds.

\section{Anisotropies in the BATSE Sample}\label{sec3}

The first indirect observational proof for the cosmological origin of GRBs was given by ~\citet {Meegan1992}. There was no concentration in the sky positions of the observed BATSE's GRBs with respect to the Galactical plane.
This indirect support of the cosmological origin was then collected and formulated by ~\citet{Tegmark1996}. This study did not find any 
concentration toward the Galactical plane and did not find also any deviation from the isotropic celestial distribution. 

\begin{figure*}[t]
        \centerline{\includegraphics[width=235pt,angle=-90]{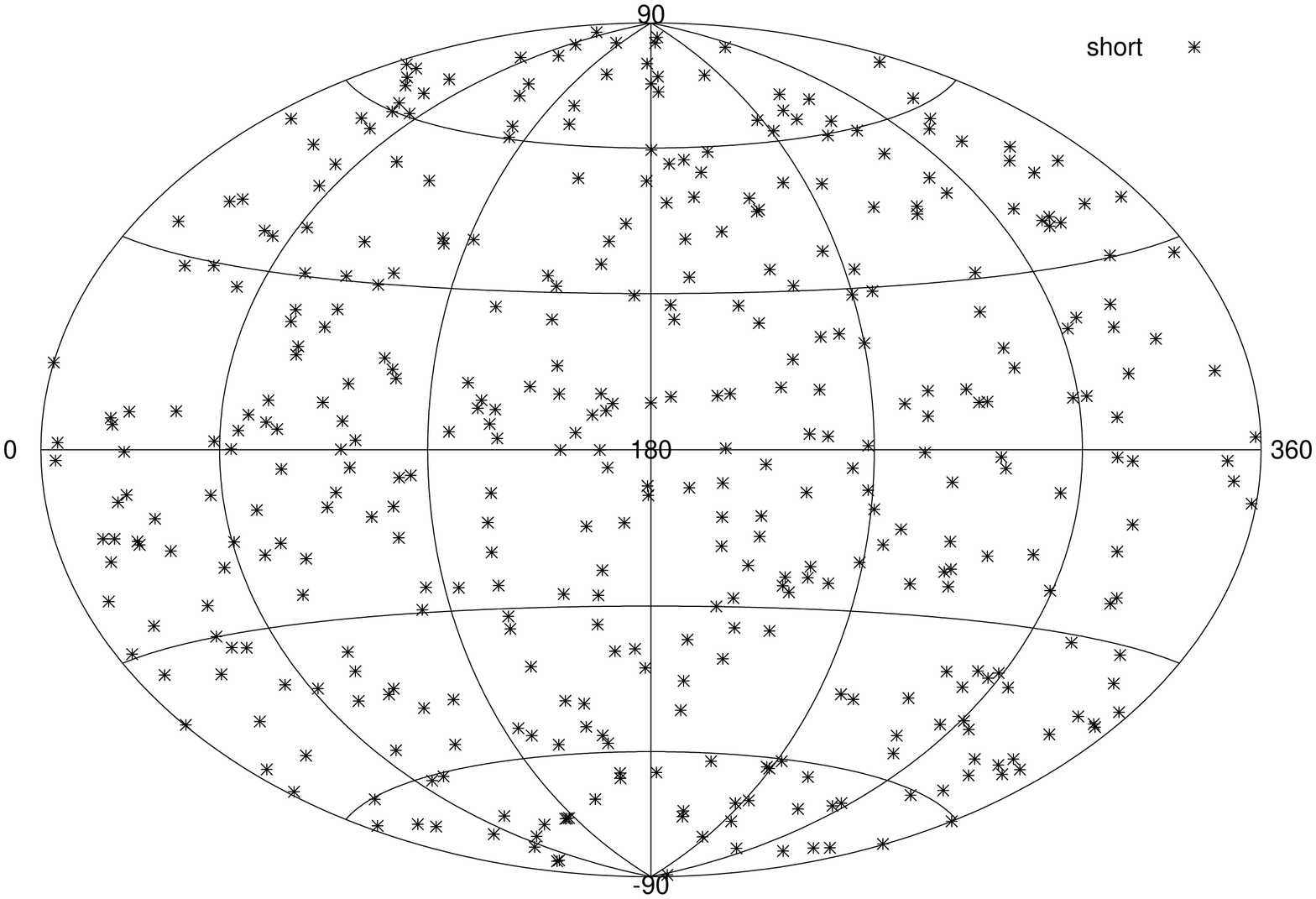}}
 \caption{The sky distribution of the BATSE's short GRBs in equatorial coordinates.\label{fig1}}
 \end{figure*}

~\citet{1998A&A...339....1B}
accepted a priori the cosmological origin of GRBs, and hence they did not search for any concentration toward the 
Galactical plane. They studied generally by statistical tests the isotropy itself in the sky distribution. Hence, in essence, they tested the fulfilment of CP, because - if fulfilled - the distribution must remain isotropic for any subclass of GRBs separately. The paper unambiguously
claims first on a high significance level that the sky distribution of the short BATSE's GRBs is not isotropic.
On  Fig.\ref{fig1}  this distribution is shown. It must be precised that the anisotropy is not given by the instrumental effects coming from the fact that the BATSE instrument did not observe uniformly at different declinations.
This proclaim about the short BATSE's GRBs was then confirmed by several other articles of the author and his collaborators 
(~\cite{1999A&AS..138..417B}, ~\cite{2000A&A...354....1M},
~\cite{Vavrek2008}). The statistical tests in  ~\citet{1998A&A...339....1B}, 
~\citet{1999A&AS..138..417B} and ~\citet{2000A&A...354....1M} used the method of spherical harmonics, ~\citet{Vavrek2008} used the minimal spanning tree, the multifractal spectra and the Voronnoi tesselation methods together with Monte Carlo simulations.

\begin{figure*}[t]
         \centerline{\includegraphics[width=235pt,angle=-90]{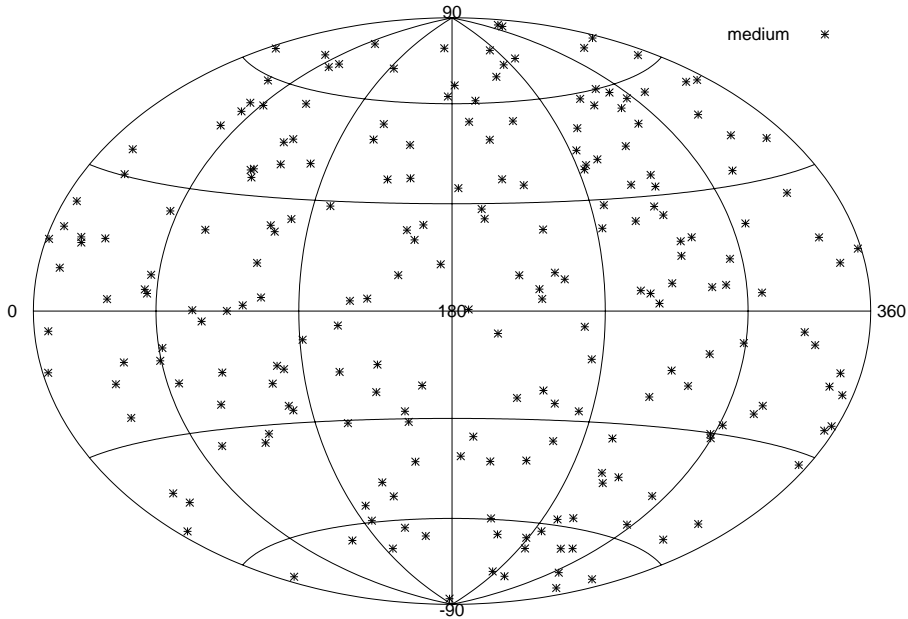}}
 \caption{The sky distribution of the BATSE's intermediate GRBs in equatorial coordinates.\label{fig2}}
 \end{figure*}

Concerning the eventual BATSE's intermediate subclass it was found that it is also distributed also anisotropically.
(~\cite{2000ApJ...539...98M}). The count-in-cell method was used.
The distribution is shown on Fig.\ref{fig2}.

\begin{figure*}[t]
         \centerline{\includegraphics[width=235pt,angle=-90]{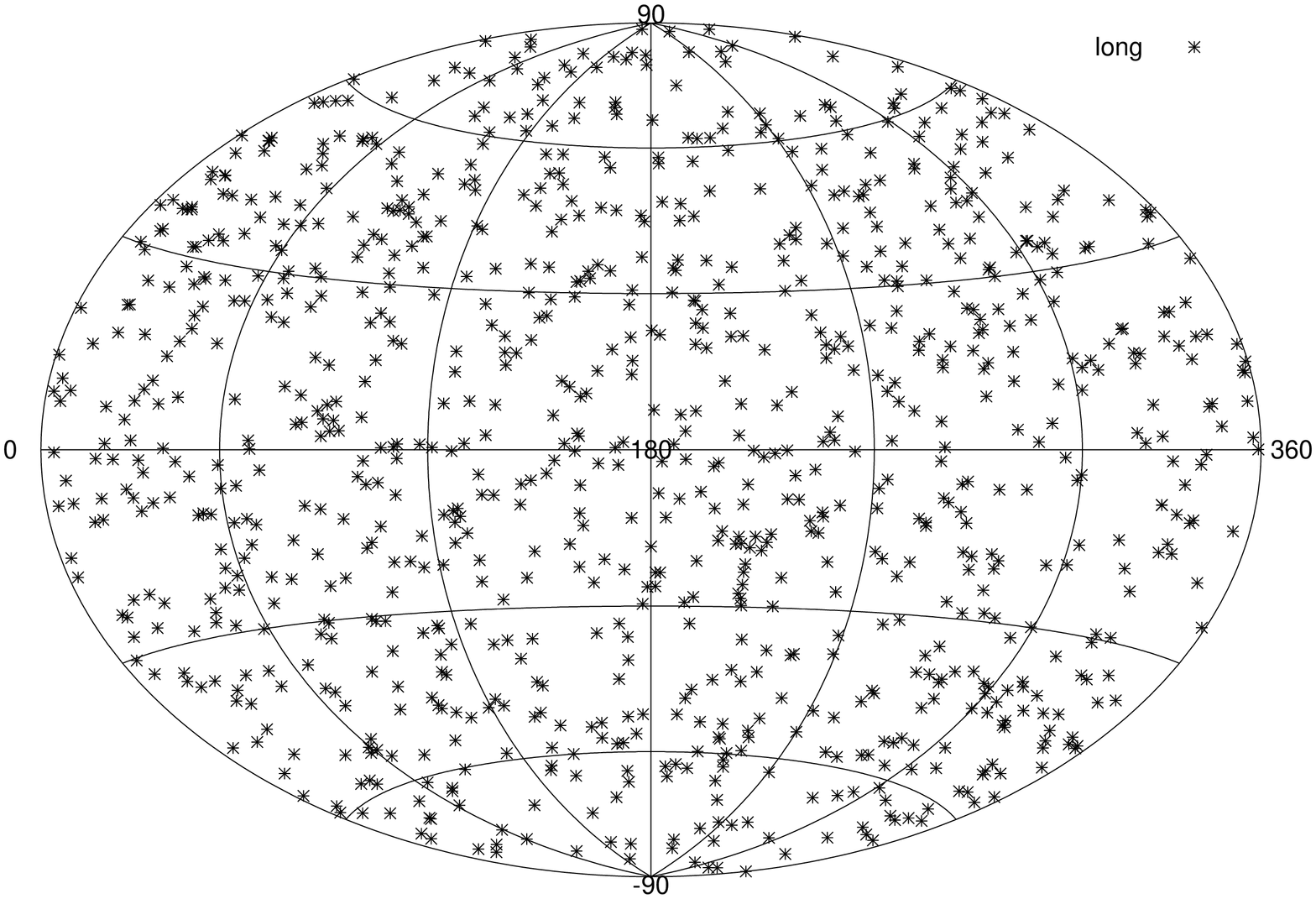}}
 \caption{The sky distribution of the BATSE's long GRBs in equatorial coordinates.\label{fig3}}
 \end{figure*}

Concerning the long BATSE's GRBs ~\citet{2003A&A...403..443M}
found the distribution to be anisotropic from the nearest-neighbourhood test. On the other hand, statistical tests from 
~\citet{Vavrek2008} gave controversial results (see Fig.\ref{fig3}). 

After ~\citet{Vavrek2008} the existence of the $Gpc$ structures and thus the breakdown of CP was declared by
~\citet{2009AIPC.1133..483M} and ~\citet{2009BaltA..18..293M}.

All these 2D studies were based on the BATSE data,
because in the BATSE dataset only few GRBs have directly measured redshifts
(see, for example, Table 1 of ~\citet{Bagoly2003}), and hence, 3D tests are not possible with the BATSE data.

\section{Other 2D statistical studies}\label{sec4}

~\citet{Ripa2017} tested the isotropy of the observed properties of GRBs in the 
whole Fermi/GBM catalog. It was studied a possibility that at different directions GRBs had different properties such as their durations, fluences, and peak fluxes at various energy bands and timescales. In other words, not the positions of GRBs were tested but their observed astrophysical properties. Later, their method - based on numerical simulations - was applied on an updated Fermi/GBM catalog, too, and extended also to the datasets of BATSE and Swift (~\cite{Ripa2019}). The observed properties of GRBs in all used datasets were found to be consistent with isotropy.
This results suggests observationally that physically GRBs should be identical in any directions, and only the positions do not show an isotropy.

~\citet{2018AstBu..73..111S} studied the grouping of galaxies around GRB021004 at $z \simeq0.56$. The method is an original combination of optical astronomy with the topic of GRB. A possibility of a galaxy cluster was found on a $3^o \times 3^o$ region. This means that one single GRB should be at a region with an excess of the number of galaxies. 

\section{3D Statistical Studies}\label{sec5}

Only a small fraction of detected GRBs at the gamma-band has measured redshifts
from the so called afterglows (either in X-rays, UV, optical, infrared and even at radio band). For these limited samples the three-dimensional statistical studies are possible.

\begin{figure*}[t]
 \centerline{\includegraphics[width=1.55\columnwidth,angle=0]{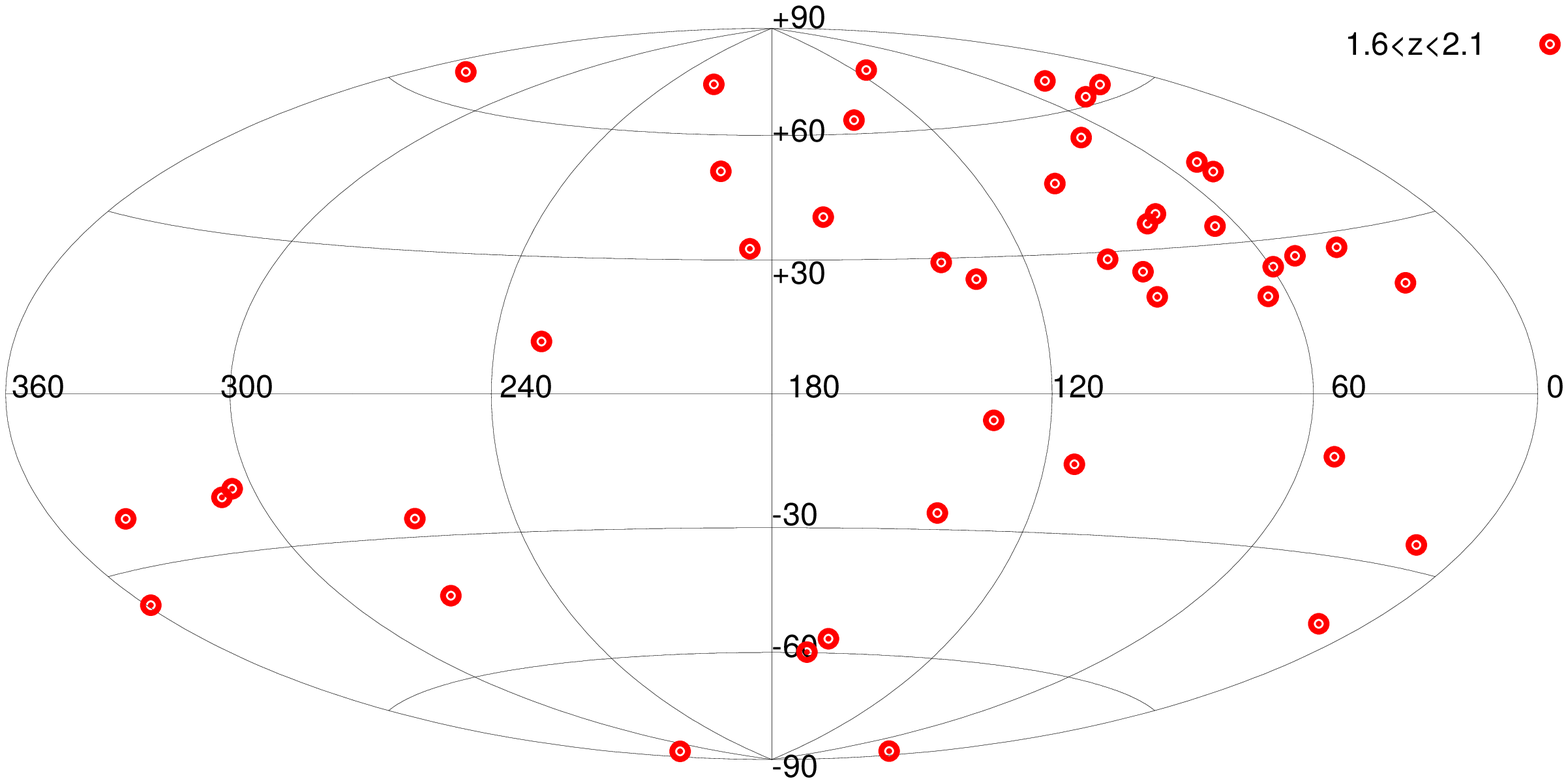}}
        \caption{The distribution of GRBs in the redshift range $1.6 < z\le 2.1$
        at Galactical coordinates.
        The cluster direction is at approx $l=88^o,b=63^o$. 
        }
\label{fig4}
\end{figure*}

Using 361 GRBs with measured optical afterglows and redshifts,
~\citet{2014A&A...561L..12H} identified a large clustering of 
GRBs at redshift $z \approx 2$ in the general direction of the constellations
of Hercules and Corona Borealis.  This excess cannot be entirely
attributed to known selection biases, making its existence due to chance
unlikely. The scale on which the clustering occurs is
about $\sim (2-3)$ $Gpc$. 
The underlying distribution of matter - suggested by this cluster - 
is again big enough to challenge (similarly to the 2D studies mentioned earlier) the standard assumption about the homogeneity and isotropy of the Universe. The position of
the structure is shown on Fig.\ref{fig4}.
The $\chi^2$ probability that this clustering is random is
$p=0.051$.

\begin{figure*}[t]
\centerline{\includegraphics[width=1.55\columnwidth,angle=0]{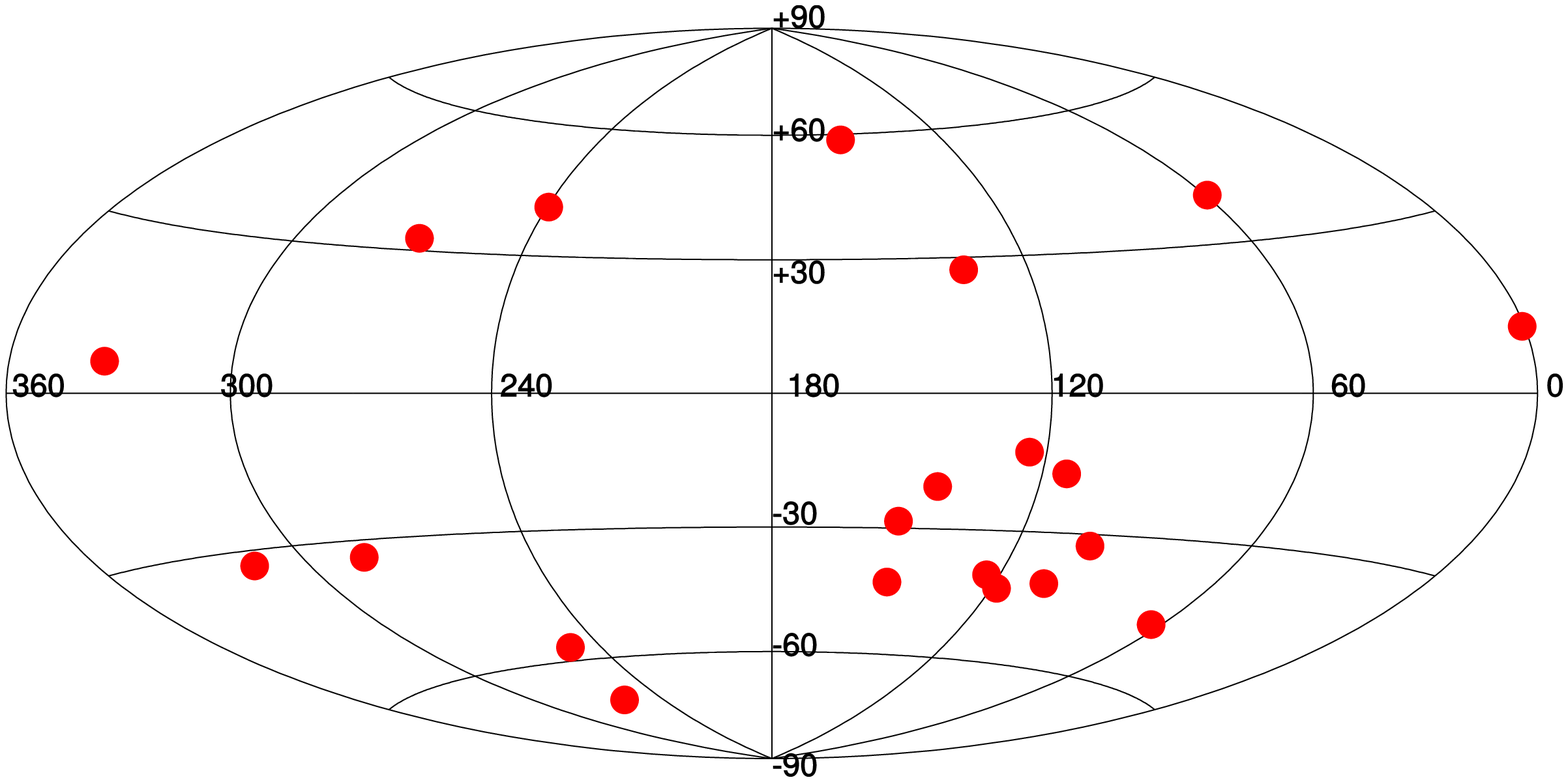}}
        \caption{
        Angular distribution of GRBs in Galactical coordinates..  
        The ring-like structure of objects is in the lower right part of the frames.
        }
\label{fig5}
\end{figure*}

~\citet{2015MNRAS.452.2236B}
(see also ~\citet{2018MNRAS.473.3169B})
was motivated by a large GRB
cluster, and the $k$-th nearest neighbour in the sample was analysed. During the analysis a large regular formation of GRBs was found: the ring is
displayed by 9 GRBs with an angular major/minor diameter of $43^o \times 30^o$ in the $0.78 < z < 0.86$ redshift 
range, and with a probability of $2\times 10^{-6}$ of being the result of a random fluctuation only (see Fig.\ref{fig5}). 

Using the same data ~\citet{2015IAUGA..2257471B}  (see also ~\citet{2016IAUS..319....3B})
reconstructed the empirical sky exposure function with the empirical radial distribution and calculated the general 3D spatial two-point correlation function of the GRBs. Signals of both the large GRB cluster and the ring  were identified. There was a third anomaly, 
caused GRB020819B and GRB050803, at a cosmologically low distance $\approx 56 \, Mpc$ from each other, with a low probability of $0.00996$ being a random fluctuation. 

\section{Conclusion}\label{sec7}

It can be claimed that the found $\sim \, Gpc$ structures of the spatial distribution of GRBs
do not allow an averaging. This results, obtained exclusively from the
observations, is at a strong contradiction with CP requiring a transition scale of homogeneity below the $\sim \, Gpc$ scale.

\bibliography{Wiley-ASNA}%

\end{document}